\documentclass[sigconf]{acmart}

\usepackage{subcaption}
\usepackage[capitalise]{cleveref}
\usepackage{tabularx}

\newcolumntype{L}{>{\raggedright\arraybackslash}X}

\copyrightyear{2024}
\acmYear{2024}
\setcopyright{rightsretained}
\acmConference[ICPE '24]{Proceedings of the 15th ACM/SPEC International Conference on Performance Engineering}{May 7--11, 2024}{London, United Kingdom}
\acmBooktitle{Proceedings of the 15th ACM/SPEC International Conference on Performance Engineering (ICPE '24), May 7--11, 2024, London, United Kingdom}\acmDOI{10.1145/3629526.3645036}
\acmISBN{979-8-4007-0444-4/24/05}

\makeatletter
\gdef\@copyrightpermission{
	\begin{minipage}{0.3\columnwidth}
		\href{https://creativecommons.org/licenses/by/4.0/}{\includegraphics[width=0.90\textwidth]{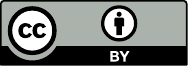}}
	\end{minipage}\hfill
	\begin{minipage}{0.7\columnwidth}
		\href{https://creativecommons.org/licenses/by/4.0/}{This work is licensed under a Creative Commons Attribution International 4.0 License.}
	\end{minipage}
	\vspace{5pt}
}
\makeatother

\begin{document}

\title{ShuffleBench: A Benchmark for Large-Scale Data Shuffling Operations with Distributed Stream Processing Frameworks}

\author{Sören Henning}
\orcid{0000-0001-6912-2549}
\affiliation{%
  \institution{JKU/Dynatrace Co-Innovation Lab,\\Johannes Kepler University Linz}%
  \city{Linz}%
  \country{Austria}%
}
\email{soeren.henning@jku.at}

\author{Adriano Vogel}
\orcid{0000-0003-3299-2641}
\affiliation{%
    \institution{JKU/Dynatrace Co-Innovation Lab,\\Johannes Kepler University Linz}%
    \city{Linz}%
    \country{Austria}%
}
\email{adriano.vogel@jku.at}

\author{Michael Leichtfried}
\orcid{0000-0002-4415-6694}
\affiliation{%
    \institution{Dynatrace Research,\\Dynatrace LLC}%
    \city{Linz}%
    \country{Austria}%
}
\email{michael.leichtfried@dynatrace.com}

\author{Otmar Ertl}
\orcid{0000-0001-7322-6332}
\affiliation{%
    \institution{Dynatrace Research,\\Dynatrace LLC}%
    \city{Linz}%
    \country{Austria}%
}
\email{otmar.ertl@dynatrace.com}

\author{Rick Rabiser}
\orcid{0000-0003-3862-1112}
\affiliation{%
    \institution{LIT CPS Lab,\\Johannes Kepler University Linz}%
    \city{Linz}%
    \country{Austria}%
}
\email{rick.rabiser@jku.at}

\begin{abstract}
  Distributed stream processing frameworks help building scalable and reliable applications that perform transformations and aggregations on continuous data streams.
  This paper introduces \emph{ShuffleBench}, a novel benchmark to evaluate the performance of modern stream processing frameworks.
  In contrast to other benchmarks, it focuses on use cases where stream processing frameworks are mainly employed for \emph{shuffling} (i.e., re-distributing) data records to perform state-local aggregations, while the actual aggregation logic is considered as black-box software components.
  ShuffleBench is inspired by requirements for near real-time analytics of a large cloud observability platform and takes up benchmarking metrics and methods for latency, throughput, and scalability established in the performance engineering research community. Although inspired by a real-world observability use case, it is highly configurable to allow domain-independent evaluations.
  ShuffleBench comes as a ready-to-use open-source software utilizing existing Kubernetes tooling and providing implementations for four state-of-the-art frameworks.
  Therefore, we expect ShuffleBench to be a valuable contribution to both industrial practitioners building stream processing applications and researchers working on new stream processing approaches.
  We complement this paper with an experimental performance evaluation that employs ShuffleBench with various configurations on Flink, Hazelcast, Kafka Streams, and Spark in a cloud-native environment. Our results show that Flink achieves the highest throughput while Hazelcast processes data streams with the lowest latency.
\end{abstract}

\ccsdesc[500]{Software and its engineering~Software performance}
\ccsdesc[500]{Computer systems organization~Cloud computing}
\ccsdesc[500]{Information systems~Stream management}

\keywords{benchmarking, data shuffling, performance, stream processing}

\maketitle

\section{Introduction}

State-of-the-art distributed stream processing frameworks such as Spark~\cite{Zaharia2016,Armbrust2018}, Flink~\cite{Carbone2015}, Kafka Streams~\cite{Sax2018,Wang2021}, or Hazelcast with its Jet engine~\cite{Gencer2021} have gained widespread adoption over the last years not only for building data analytics pipelines, but also for implementing core business logic in software-based organizations~\cite{Katsifodimos2019,Fragkoulis2023}.
Such frameworks support software engineers in building highly scalable, reliable, and efficient applications that process continuous data streams of massive volume.
They provide high-level APIs and domain-specific languages to define the processing logic as directed acyclic processing graphs that filter, transform, aggregate, and merge data streams.

Over the last decade, several works have been published that evaluate the performance of distributed stream processing frameworks or propose new benchmarks, evaluation methods, and auxiliary tools~\cite{Karimov2018,vanDongen2020,Bordin2020,Hesse2021,BDR2021,SEAA2023}.
Available benchmarks usually contain one or a few task samples that fulfill domain-specific use cases, for example, for analyzing car traffic data~\cite{Arasu2004,vanDongen2020} or aggregating Industrial Internet of Things (IIoT) sensor measurements~\cite{BDR2021,Hesse2021}.
To implement these task samples (also referred to as queries), the benchmarks define the domain-specific processing logic with the frameworks' high-level APIs.

In this paper, we study a different, more general use case, where the actual domain-specific processing logic is out of the scope of a stream processing framework, but stream processing frameworks are still used for their abstractions for cluster management, means to scale out the data processing, fault-tolerant state management, well-defined processing guarantees (e.g., exactly-once or at-least-once), and rich ecosystem of documentation, support, and associated tooling.
We illustrate this use case inspired by requirements of a large cloud observability platform, where potentially thousands or millions of stateful black-box software components have to receive and process selected data records.
The literature is currently missing a well-defined evaluation method for this use case and previous work found that the performance of stream processing frameworks highly depends on the use case~\cite{JSS2024}.

Therefore, we propose \emph{ShuffleBench} as a new stream processing benchmark focusing on the use of stream processing frameworks for \emph{shuffling} (i.e., re-partitioning) data streams to efficiently process data in high numbers of stateful components.
ShuffleBench provides well-defined metrics, measurement methods, and a highly configurable task sample.
Thus, it allows researchers and practitioners to evaluate stream processing frameworks with respect to performance attributes such as throughput, latency, and scalability and regarding a representative use case.
Our research design combines requirements identified from a large cloud observability platform with established benchmarking metrics, methods, and techniques from the performance engineering research community.

\paragraph{Contributions} In summary, we make the following contributions with this paper to industry practitioners and the research community:
\begin{itemize}
    \item We propose ShuffleBench, a new benchmark for distributed stream processing frameworks. It addresses a use case not covered by existing benchmarks and is highly configurable for different assorted characteristics.
    \item We provide open-source implementations\footnote{\url{https://github.com/dynatrace-research/ShuffleBench}} of ShuffleBench for different state-of-the-art stream processing frameworks as well as associated tooling to automate running benchmarks in Kubernetes-based cloud environments.
    \item We conduct an experimental evaluation covering throughput and latency that are highly relevant metrics for stream processing~\cite{SEAA2023}. Our evaluation covers the open-source stream processing frameworks Flink, Hazelcast, Kafka Streams, and Spark Structured Streaming due to their industry acceptance and academic relevance~\cite{SEAA2023}.
    We provide a replication package and the collected data of our experiments as supplemental material~\cite{ReplicationPackage}, such that other researchers and practitioners may repeat and extend our work.
\end{itemize}

\paragraph{Outline}
The remainder of this paper is structured as follows:
\Cref{sec:background} introduces the fundamental concepts of distributed stream processing frameworks and related benchmarking studies.
\Cref{sec:use-case} describes the use of stream processing for large-scale data shuffling illustrated by industrial requirements.
Afterward, \cref{sec:benchmark} presents ShuffleBench, our proposal for a new stream processing benchmark.
\Cref{sec:evaluation} employs ShuffleBench to evaluate four stream processing frameworks regarding their throughput and latency for different configurations.
\Cref{sec:conclusions} concludes this work and discusses future work.

\section{Background and Related Work}\label{sec:background}

In the following, we briefly introduce the fundamental concepts of modern stream processing frameworks and discuss related work on benchmarking stream processing frameworks.

\subsection{Distributed Stream Processing}\label{sec:background:stream-processing}

Stream processing frameworks perform operations such as filterings, transformations, or aggregations in near-real time on continuous streams of data~\cite{Hirzel2013}.
State-of-the-art frameworks are designed for high throughput and low-latency processing, while also scaling with massive amounts of data~\cite{Fragkoulis2023,JSS2024}.
To address these requirements, they run in a distributed fashion on commodity hardware. A key advantage of stream processing frameworks is that they provide dataflow models that abstract aspects such as cluster management, state management, and time semantics from their users~\cite{Akidau2015,Sax2018}.
With such models, engineers describe the processing logic in directed acyclic dataflow graphs of processing operators.
The frameworks allow the initiation of multiple instances across various compute nodes, containers, or with multiple threads, with each instance handling a distinct portion of the data.
While the isolated processing of data records remains unaffected by the assignment of data portions to instances, processing that depends on previous data records, such as  aggregations, requires state management.
Similar to the MapReduce~\cite{Dean2008} programming model, keys are assigned to records before a stateful operation. This allows the stream processing frameworks routing all records with the same key to the same instance.
Consequently, no state synchronization among instances is required, which allows operating stream processing applications at significantly lower costs compared to, for example, Functions-as-a-Service~\cite{Pfandzelter2022}.
When a processing operator modifies the key of a record and a subsequent operator performs a stateful operation, the framework divides the dataflow graph into subgraphs that can be independently processed by different instances.

Popular stream processing frameworks include Apache Flink~\cite{Carbone2015}, Hazelcast with its Jet engine~\cite{Gencer2021}, Apache Kafka Streams~\cite{Sax2018,Wang2021}, and Apache Spark~\cite{Zaharia2016} with its Structured Streaming engine~\cite{Armbrust2018}.
Although all these frameworks follow similar concepts, several differences in their design decisions, programming functionalities, and execution models can be noted.
Whereas Flink and Spark follow a master-worker cluster architecture, Hazelcast and Kafka Streams instances form clusters and perform the necessary coordination internally.
Moreover, Hazelcast and Kafka Streams can be embedded as library into applications, while Flink and Spark are the applications themselves.
Spark is different from the other frameworks as it processes the data streams as a sequence of batches (called micro-batches), whereas the other frameworks process record by record as they arrive.
Hazelcast's Jet engine is special by being built on top of the Hazelcast IMDG distributed, in-memory object store. It differs in its execution model, which is based on a concept similar to coroutines and cooperative threads to process data at very low latency~\cite{Gencer2021}.
Kafka Streams tightly integrates with the Apache Kafka messaging system~\cite{Kreps2011}. It does not support other data sources and sinks than Kafka, but benefits from lightweight coordination based on Kafka's consumer group protocol.

Stream processing frameworks often read data from or write data to messaging systems.
Such messaging systems serve both as a scalable middleware between different systems and services as well as the necessary infrastructure to ensure fault tolerance. For this purpose, industry-grade messaging systems such as Apache Kafka~\cite{Kreps2011,Wang2021} employ an immutable, sequentially appended log structure to store and replicate records across distributed nodes.

\begin{table*}
    \caption{Overview of related open-source stream processing benchmarks and associated benchmarking studies}
    \label{tab:related-work}
    \begin{tabularx}{\textwidth}{lLLLLL}
        \toprule
        & DSPBench~\cite{Bordin2020} & OSPBench~\cite{vanDongen2020,vanDongen2021a,vanDongen2021b} & ESPBench~\cite{Hesse2021} & Theodolite~\cite{BDR2021,JSS2024} & \textbf{ShuffleBench (This work)}\\
        \midrule
        Benchmark inspiration & From the literature & Car traffic & IIoT & IIoT & Cloud observability \\
        Application metrics & Throughput, latency & Throughput, latency & Latency & Scalability & Throughput, latency\\
        Frameworks & Spark, Storm & Flink, Kafka Streams, Spark, Spark Structured Streaming & Beam (Flink, Hazelcast, Spark) & Beam, Flink, Hazelcast, Kafka Streams & Flink, Hazelcast, Kafka Streams, Spark Structured Streaming\\
        Deployment & Virtual machines & Containers, DC/OS & Virtual machines & Containers, Kubernetes &  Containers, Kubernetes\\
        Data size & 1 to 60 KB & approx. 200 bytes & up to 280 bytes & up to 100 bytes & configurable \\
        Processed record/second & up to 100k & up to 900k & up to 10k & up to one million &  up to one million \\
        Customizable state size & No & No & No & No & Yes\\
        \bottomrule
    \end{tabularx}
\end{table*}

\subsection{Benchmarking Stream Processing Frameworks}\label{sec:background:benchmarking}

A considerable number of studies proposing new benchmarks or reporting on experimental evaluations with available benchmarks have been published over the last years~\cite{SEAA2023}.
Recent examples of open-source benchmarks include DSPBench~\cite{Bordin2020}, OSPBench~\cite{vanDongen2020}, ESPBench~\cite{Hesse2021}, or the streaming benchmarks from the Theodolite framework~\cite{BDR2021}.
\cref{tab:related-work} provided an overview of these benchmarks and a comparison with ShuffleBench.
For a systematic and comprehensive review of the literature on stream processing benchmarking, we refer to our recent studies~\cite{SEAA2023,JSS2024}.

The most important difference between other benchmarks and ShuffleBench is the addressed use case. Most benchmarks re-sample domain-specific analytics applications, for example, for IIoT data streams~\cite{Hesse2021,BDR2021}, car traffic data~\cite{vanDongen2020}, or online gaming~\cite{Karimov2018}. Our paper in contrast investigates a more generic use case, which addresses a specific type of software architecture instead of a specific domain.

Despite the addressed use case, we found that benchmarks proposed by the research community are often not applicable for industry-grade performance evaluations.
They are often not available as open-source software or not actively maintained, lack automation, and do not use deployments that are representative for production systems. For example, although it is nowadays common to run stream processing applications in containerized cloud-native environments, there is only one benchmark besides ShuffleBench specifically tailored to such deployments~\cite{BDR2021}.
Likewise, results of performance evaluations conducted with these benchmarks are often not transferable to industry use cases, for example, because they evaluate frameworks only at small-scale deployments~\cite{Bordin2020}. Other benchmarks or evaluations do not define metrics and measurement methods, which makes the research difficult to reproduce and extend.

Nevertheless, the literature provides valuable research on evaluation metrics and measurement methods~\cite{Karimov2018,vanDongen2020,LTB2021}. We build upon such research to provide well-defined metrics and measurement methods with ShuffleBench.

\section{Large-Scale Data Shuffling}\label{sec:use-case}

We illustrate the use case of large-scale data shuffling with requirements for continuous dashboard queries and real-time alerting of a market-leading cloud observability platform. Using a powerful query language, it allows internal and external clients to define complex rules to aggregate and correlate different data sources such as metrics, events, logs, and traces.
From a software architecture perspective, each registered query can be considered as a runtime software component, which continuously receives all data records that are affected by this query. We call these components \emph{real-time consumers}. They have to manage state across multiple input records and might produce outputs.
For example, a consumer that performs an anomaly detection by correlating logs and performance metrics might produce an output event when it detects an anomaly.

A core requirement for a corresponding query runtime is to efficiently route data to the respective consumers while also having cluster management abstractions, means to scale out the data processing, fault-tolerant state management, and well-defined processing guarantees.
State-of-the-art stream processing frameworks fulfill these properties.
They also provide programmatic APIs or dedicated SQL-like languages to define complex queries on data streams (see related benchmarks in \cref{sec:background}).
However, those only have limited relevance for our use case as we are facing a high amount of queries that have to be executed in parallel, although each query only requires a very small portion of the overall data volume. Therefore, it is not required to parallelize or even distribute the execution of a single query but only to route those data records to a consumer as needed.
Moreover, considering that dedicated software components (e.g., anomaly detection models) can include the logic for operations such as joins or sliding windows, these features might not be required by the stream processing framework.

In addition to the observability use case described here, we expect other real-world systems to have similar requirements. For example, the stateful Function-as-a-Service~\cite{Akhter2019} runtime Apache Flink Stateful Functions\footnote{\url{https://nightlies.apache.org/flink/flink-statefun-docs-master/}} is used in different contexts, but based on a very similar architecture as our benchmark (see the following section).

\section{The ShuffleBench Benchmark}\label{sec:benchmark}

We combine the architectural requirements described above with best practices of the performance engineering community and industrial consortia~\cite{Kistowski2015, Kounev2020,Papadopoulos2021,Hasselbring2021}.

\subsection{Benchmark Design}
We base our benchmark design on the ACM SIGSOFT Empirical Standard for benchmarking~\cite{Ralph2021,Hasselbring2021}.\footnote{\url{https://acmsigsoft.github.io/EmpiricalStandards/docs/?standard=Benchmarking}} It distinguishes between the following four components of a benchmark: the qualities to be evaluated, metrics to quantify these qualities, measurement methods for these metrics, and task samples to be evaluated with the measurement methods. As discussed in \cref{sec:background}, the task samples of existing benchmarks have several shortcomings, making them impractical to study our described use case. On the other hand, several evaluation metrics, methods, and tools introduced with other benchmarks are applicable independently of the specific use case.
According to the Empirical Standard, we, therefore, propose a new benchmark, which, on the one hand, introduces a new task sample but, on the other hand, takes up existing metrics and measurement methods that have demonstrated their effectiveness in the literature.
This way, we address benchmark quality attributes such as relevance, reproducibility, fairness, verifiability, and usability as required by industrial consortia and the research community~\cite{Kistowski2015}.

\begin{figure}
    \centering
    \includegraphics[width=\linewidth]{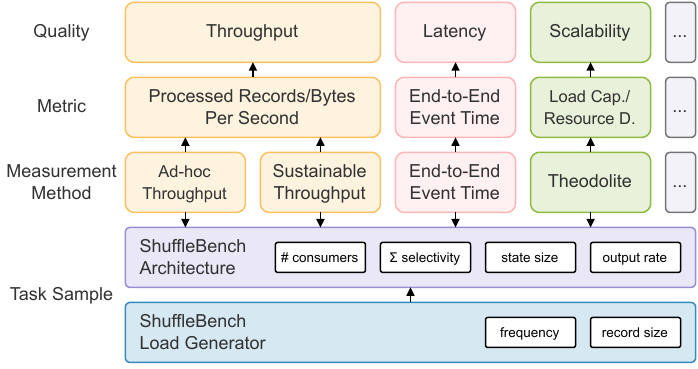}
    \caption{Overview of \emph{ShuffleBench} benchmark components according to the Empirical Standard for benchmarking.}
    \label{fig:components}
\end{figure}

\Cref{fig:components} provides an overview of our benchmark's components.
In the following, we first describe the task sample in terms of its dataflow architecture to be implemented by different stream processing frameworks and a corresponding load generator.
Afterward, we describe the qualities, metrics, and measurement methods to assess the task sample implementations.

\subsection{The ShuffleBench Task Sample}

The ShuffleBench task sample consists of a dataflow architecture to be implemented by different stream processing frameworks and a load generator to stress the framework. Both are highly configurable to allow for evaluations in different scenarios. 

\subsubsection{Benchmark Dataflow Architecture}

\begin{figure*}
    \centering
    \includegraphics[width=\linewidth]{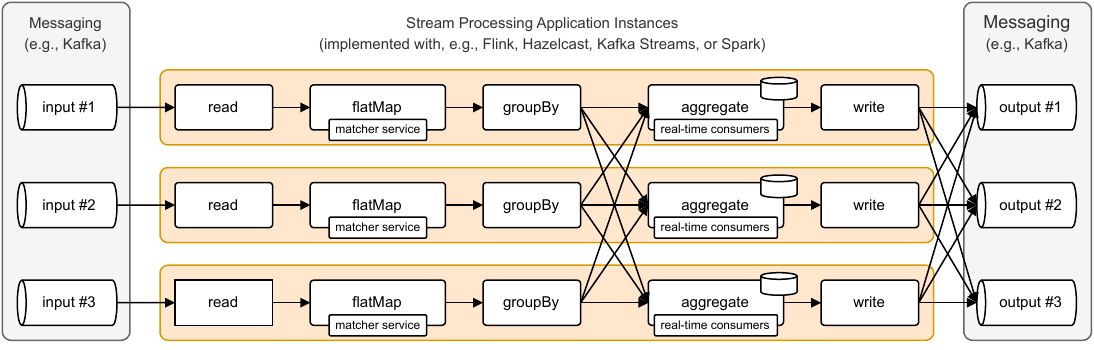}
    \caption{The \emph{ShuffleBench} dataflow architecture at runtime for three stream processing application instances.}
    \label{fig:architecture}
\end{figure*}

A well-suited way to meet the described use case is a MapReduce-like architecture \cite{Dean2008} on continuous data streams as it can be built with modern stream processing frameworks.
\Cref{fig:architecture} depicts our benchmark architecture for a corresponding stream processing application.
It represents the long-running application as a static dataflow graph, with data moving along the edges.
It can be deployed with multiple instances, which execute the same processing logic but on different data subsets.
The data processing starts by reading data records from a messaging system such as Kafka.
Kafka topics are partitioned, allowing each instance of the stream processing application to subscribe to a dedicated set of partitions.
In general, we cannot assume any specific partitioning on the data.
After ingestion, the \emph{matcher service} finds the relevant queries for each record.
The matcher logic is wrapped in a \emph{flatMap} operation of a stream processing framework that duplicates an incoming record for each relevant query while assigning a query-identifying key to each duplicate.
In a subsequent operation, the data is re-partitioned among all instances such that all records with the same query key are forwarded to the same instance.
This is done with an operation of the stream processing framework often called \emph{groupBy}.
In the next step, the actual black-box query logic is executed, which is stateful by aggregating multiple records.
We abstract the query logic in \emph{real-time consumers}, wrapped in an \emph{aggregate} operation of the stream processing framework that manages the state.
These real-time consumers adhere to a simple interface: they consume an incoming record and the previous state and output the updated state and, optionally, an event.
Finally, these output events are written to another Kafka topic.

\subsubsection{Benchmark Implementations}

We implemented the proposed benchmark for the four stream processing frameworks Flink, Hazelcast, Kafka Streams, and Spark Structured Streaming. To provide a fair comparison, the implementation of the matcher service and the real-time consumers (which would be domain-specific in production) are shared among all frameworks.
The matcher service is configured with a set of rules, which define their selectivity, i.e., the probability that this rule matches a record.
The actual stateful aggregations logic is currently the same for each query: Every incoming record is updates to the state of the respective real-time consumer. The state size is configurable, but always includes the count of received records, a checksum, and the associated timestamps of the first and the last record (see \cref{sec:metrics:latency}). Real-time consumers emit an output event if the count is divisible by a configured value to simulate something like a generated alert. 

\subsubsection{Load Generator}

To simulate incoming observability data, we provide a load generator that creates data records at a configurable frequency with random byte content of configurable size.
We decided to use this way of generating data instead of replaying historical data to make performance evaluations more reproducible and ShuffleBench applicable to other domains~\cite{Kistowski2015}.
The load generator can be deployed in a distributed fashion and writes the generated records to Kafka. 

\subsubsection{Benchmark Configuration Options}
Our ShuffleBench implementations are highly configurable to evaluate frameworks for different use cases of large-scale data shuffling tasks. This includes the size of incoming records, the number of different real-time consumers, the total selectivity for all real-time consumers, the distribution of individual selectivities, the real-time consumer's state size, and their output frequency. Additionally, all stream processing frameworks have a wide range of configuration and deployment options that potentially impact throughput, latency, scalability, and fault tolerance. Many of them can easily be set with ShuffleBench to support experimental comparisons.

\subsection{Qualities, Metrics and Measurement Methods}

According to requirements from industry that motivated us to design ShuffleBench, we currently support benchmarking the qualities throughput, latency, and scalability with ShuffleBench. In the following, we describe our employed metrics and measurement methods for all qualities in detail. These descriptions, along with our executable benchmarking software (see \cref{sec:benchmark:bundle}) support reproducibility of benchmarking studies conducted with ShuffleBench, such as our evaluation in \cref{sec:evaluation}. %

\subsubsection{Throughput}\label{sec:metrics:throughput}

We support measuring the throughput of stream processing frameworks in terms of the number of incoming records that can be processed per second. Note that the throughput in terms of processed bytes per second can be directly derived from that.
Assuming validated functional correctness, all frameworks process each record exactly once in the absence of failures, which means that the output throughput is proportional to the input throughput. %

Essentially, two measurement methods for throughput can be observed in the literature. We refer to them as \emph{ad-hoc throughput} and \emph{sustainable throughput}. As both have their pros and cons, we support both within ShuffleBench and evaluate them in \cref{sec:evaluation}.

Both measurement methods have in common that they only monitor the messaging system. This has the significant advantage that the throughput measurements do not influence the execution of the application. In fact, the measurement method is fully independent of the benchmarked framework and, thus, can also be used on arbitrary other task samples, including real-world applications.

\paragraph{Ad-hoc Throughput} %

Ad-hoc throughput measurements can be performed by generating and sending a constant number of records per second to the messaging system and measuring how much of these can be processed. It has to be ensured that the generated data volume is at least as high as the processing rate of the stream processing framework to not limit it.
Several studies appear to have conducted throughput measurements using this approach \cite{Karakaya2017,Nasiri2019}. %
In ShuffleBench, ad-hoc throughput is measured by tracking the rate of committed offsets at the messaging system~\cite{Kreps2011}.

The key advantage of the ad-hoc throughput method is that it can be performed in a short time.
However, the obtained throughput results might not fully reflect the real behavior of the stream processing framework when being subject to that load~\cite{LTB2021}.
For example, optimizations such as batching might allow a system to ingest data at a higher rate when reading from a large backlog compared to when data is ingested as it is generated.

\paragraph{Sustainable Throughput}

Sustainable throughput~\cite{Karimov2018} is defined as the maximum load a system can sustain without violating performance goals. %
Such a performance goal can be a limit on the event latency~\cite{Karimov2018,vanDongen2020} or the maximum tolerable increase in the number of queued messages~\cite{vanDongen2020,BDR2021,JSS2024}.
In ShuffleBench, sustainable throughput is measured by running multiple independent experiments, in which the generated load is increased from experiment to experiment and performance goals are evaluated~\cite{LTB2021}. Per default, we evaluate whether the number of queued messages increases substantially over time. This can again be obtained by tracking available message offsets and consumed message offsets at the messaging system~\cite{Kreps2011}. However, we also allow for using custom (e.g., use case-specific) performance goals.

The sustainable throughput measurement method overcomes the limitations of the ad-hoc throughput method. It better reflects how a tested system would behave in a real-world deployment.
On the downside, however, this method can only find a range in which the real achievable throughput lies.
Moreover, it has significantly longer execution times compared to the ad-hoc method. Instead of evaluating a system with one constant load rate, experiments have to be executed for different load rates.

\subsubsection{Latency}\label{sec:metrics:latency}

There are several different notions of latency in stream processing~\cite{Karimov2018,Chandramouli2011}.
In ShuffleBench, we quantify latency from a user perspective by measuring how long it takes for an input event to generate an output event.
Although most frameworks collect some kind of latency metrics, their measurement methods differ, preventing a fair comparison.
Moreover, in contrast to throughput, we cannot measure latency purely in the messaging system since input and output events are not natively linked to each other.

In line with \citet{vanDongen2020} and \citet{Hesse2021},
we measure latency as the time difference between writing a record to the input messaging topic and the time a record is written to the output topic.
These timestamps can be retrieved in a framework-independent way as messaging systems such as Kafka assign each record the current timestamp when appending it to the log.
To correlate both timestamps, we extract the timestamp of the input record and append it to the record's payload. Depending on the framework, this is either done as part of the \emph{read} operator (see \cref{fig:architecture}) or in an intermediate stateless \emph{map} operation. In the stateful aggregation step, we use this timestamp and include it in the state. Since the whole state is emitted as part of an output event, the final data written to Kafka contains both the time the input event has been written to Kafka and the time the output event has been written to Kafka.
ShuffleBench comes with a \emph{latency exporter} that reads all output events and computes a frequency distribution of their latency. The latency exporter can be scaled to multiple instances to also cope with large output data volume.

\subsubsection{Scalability}

We adopt the \emph{Theodolite} scalability benchmarking method for stream processing systems in cloud-native environments~\cite{EMSE2022, LTB2021}. It provides two metrics for quantifying scalability: The \emph{resource demand} metric describes how the number of required computing resources evolves with increasing load, whereas the \emph{load capacity} describes how the maximum processable load evolves with increasing computing resources. The Theodolite measurement method extends the sustainable throughput method to a two-dimensional search space. It encompasses search strategies two reduce the number of independent experiments to be executed.

\subsection{Open-Source Benchmark Availability}\label{sec:benchmark:bundle}

We provide ShuffleBench as free and open-source research software to the community.
ShuffleBench utilizes and extends the benchmarking framework Theodolite~\cite{EMSE2022,IC2E2022} to automate the benchmark execution in Kubernetes-based cloud environments.
This includes the declarative definition of benchmark experiments, automated setup and teardown of all involved software components (i.e., stream processing frameworks, load generator, and the messaging system) as well as the collection of measurement data.

Our benchmark implementations for different stream processing frameworks as well as the load generator and the latency exporter are available as source code, Java archive files, container images, Kubernetes manifests, and Theodolite benchmark manifests. This allows the community to engage at different levels.

\section{Experimental Evaluation}\label{sec:evaluation}

We employ ShuffleBench to experimentally evaluate the performance of Flink (v. 1.17), Hazelcast (v. 5.3), Kafka Streams (v. 3.5), and Spark Structured Streaming (v. 3.4, in the following simply referred to as Spark). This evaluation serves two purposes: First, it compares how different stream processing frameworks compete regarding large-scale data shuffling use cases to assist in selecting the right technology. Second, it demonstrates how ShuffleBench allows researchers and practitioners to conduct their own experiments with different configurations of the benchmark.

After a brief description of our experimental setup (\cref{sec:evaluation:setup}), we conduct a set of experiments.
\Cref{sec:evaluation:throughput} starts by a baseline evaluation of throughput using both measurement methods presented in \cref{sec:metrics:throughput}.
Likewise, \cref{sec:evaluation:latency} reports on our baseline latency evaluation.
Afterward, \cref{sec:evaluation:deployment} repeats these experiments with a modified deployment, 
\cref{sec:evaluation:recordsize} with different generated record sizes,
\cref{sec:evaluation:numconsumers} with a different number of real-time consumers, and
\cref{sec:evaluation:selectivity} with different total selectivities of the matcher service.
We discuss the results in the context of related work in \cref{sec:evaluation:discussion} and threats to validity in \cref{sec:evaluation:threats-to-validity}.
We provide a replication package and the collected data of our experiments as supplemental material~\cite{ReplicationPackage}, such that other researchers and practitioners may repeat and extend our work.

\subsection{Experimental Setup}\label{sec:evaluation:setup}

We conduct our experimental evaluation in a Kubernetes cluster managed by the Elastic Kubernetes Service of Amazon Web Services.
The cluster consists of 10~nodes provisioned in the \emph{us-east-1} region: 3~\emph{m6i.xlarge} nodes run the stream processing framework, 3~\emph{m6i.2xlarge} nodes run one Kafka broker each, and 4~\emph{m6i.xlarge} nodes run the load generator instances plus additional benchmarking infrastructure.%
\footnote{As part of our replication package~\cite{ReplicationPackage}, we also provide the exact setup using established Infrastructure-as-Code tooling.}
Unless stated differently, we use the following configurations: We deploy the stream processing application with 9~application instances (3~per cluster node). Each instance is assigned 4~GB of memory and a single virtual CPU core, resulting in a total parallelism of 9.
Except for a few adjustments for better comparability, we test all frameworks with their default configurations.
We set up one million real-time consumers that all have the same selectivity, which sum up to 20\,\%, meaning that each record is forwarded to $0.2$ consumers on average.
Each consumer emits an output event for every tenth record.
Generated records have a size of 1024~bytes. %
To increase statistical rigor, we run each experiment for 15~minutes and repeat it three times. We consider this sufficient as our results in the following sections show no large deviations across repetitions.

\subsection{Baseline Evaluation of Throughput}\label{sec:evaluation:throughput}

\begin{figure*}%
	\begin{subfigure}[c]{0.33\linewidth}%
		\includegraphics[width=\linewidth]{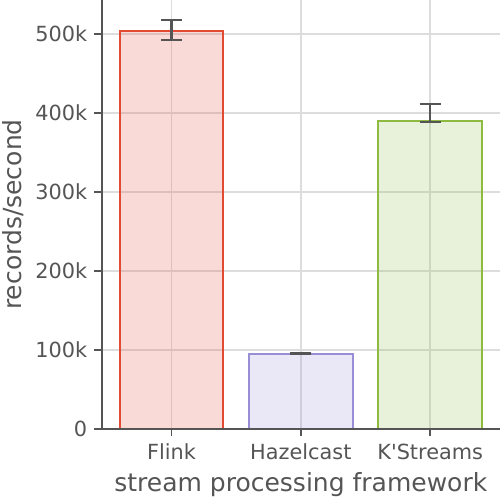}%
		\caption{Ad-hoc throughput}%
		\label{fig:barplot-baseline-atp}
	\end{subfigure}%
	\hfill%
	\begin{subfigure}[c]{0.33\linewidth}%
		\includegraphics[width=\linewidth]{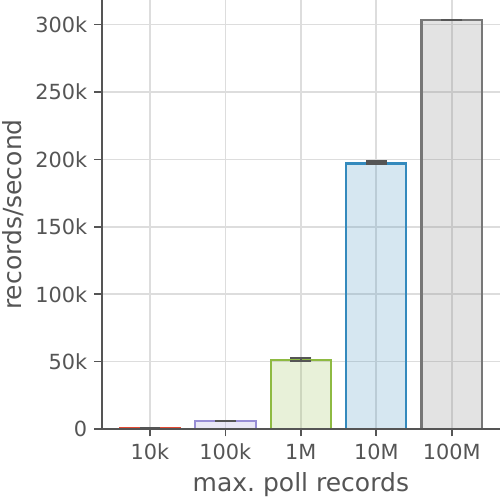}%
		\caption{Ad-hoc throughput (Spark)}%
		\label{fig:barplot-baseline-atp-spark}%
	\end{subfigure}
	\hfill%
	\begin{subfigure}[c]{0.33\linewidth}%
		\includegraphics[width=\linewidth]{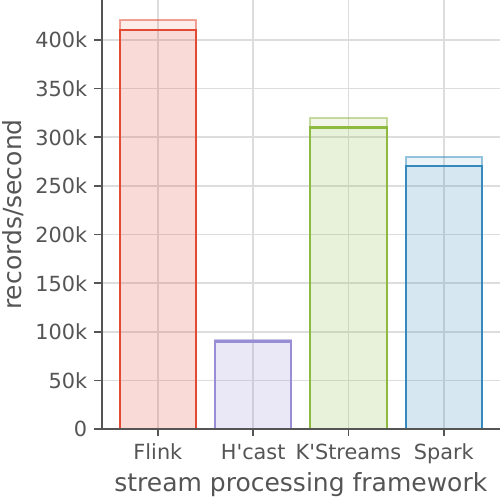}%
		\caption{Sustainable throughput}%
		\label{fig:barplot-baseline-stp}%
	\end{subfigure}
	\caption{Baseline throughput results obtained with the ad-hoc measurement method of Flink, Hazelcast, and Kafka Streams (a); of Spark for different limitations on the maximum number of pulled records per batch (b); and with the sustainable measurement method of Flink, Hazelcast, Kafka Streams, and Spark (c).}
\end{figure*}

In \cref{sec:metrics:throughput}, we discuss how the \emph{ad-hoc throughput} method provides throughput results faster than the more realistic \emph{sustainable throughput} method.
In this evaluation, we compare the results of both methods based on experiments.
For the ad-hoc throughput experiments, we generate 1~million records per second and monitor how many are processed per second by the frameworks.
For the sustainable throughput experiments, we generate records with different frequencies and determine the maximum frequency at which the number of queued records in the Kafka input topic does not substantially increase over time (the performance goal, see our previous work for a detailed explanation of this method~\cite{JSS2024}).
In the following, we first discuss the results of Flink, Kafka Streams, and Hazelcast as they allow for a straightforward interpretation, followed by a more detailed discussion of the results for Spark.

\subsubsection{Results of Flink, Hazelcast, and Kafka Streams}

\Cref{fig:barplot-baseline-atp} shows the results of the ad-hoc throughput measurement method for Flink, Hazelcast, and Kafka Streams. We can observe a clear ranking in which Flink achieves the highest throughput, followed by Kafka Streams, and a considerably lower throughput of Hazelcast. %

We contrast these results by the results of the sustainable throughput method as shown in \cref{fig:barplot-baseline-stp}.
Considering the results obtained by the sustainable throughput method as representative of a real deployment, we can see that the ad-hoc method overestimates the throughput of Flink and Kafka Streams by up to 20\%.
For Hazelcast, we can see no difference. %
Despite their overestimation, ad-hoc measurements are still useful since the ranking of frameworks is the same with both methods and ad-hoc measurements can be performed significantly faster.
For these reasons, we apply the ad-hoc method for the throughput measurements in the following evaluation.

\subsubsection{Results of Spark}

Unless further adjusted, ad-hoc throughput measurements are not meaningful in Spark, because Spark ingests all available data in a batch, processes this batch, and then ingests the next batch of all available data. Data that is generated at a larger volume than can be processed leads to ever-increasing batch sizes and, thus, to ever-increasing batch processing time. However, Spark allows constraining the maximum number of records pulled per batch. We experiment with different limits and show how they impact the achieved throughput in \cref{fig:barplot-baseline-atp-spark}. In short, the results demonstrate the intuition that pulling more records increases the throughput. However, this comes at the cost of high latency. 
For instance, pulling in batches of 10~million records achieves a throughput of approximately 200\,000 records per second, but then data is retrieved only every 50~seconds causing high latency (see \cref{sec:evaluation:latency}).

With the sustainable throughput method, we determine the maximum throughput without a persistent increase in queued messages.
As shown in \cref{fig:barplot-baseline-stp}, Spark achieves a throughput of 270\,000 to 280\,000 records per second.
However, this throughput comes again at the cost of large batches with processing times -- and thus latencies (see \cref{sec:evaluation:latency}) --
of several seconds. %
Smaller data volumes lead to reduced batch processing times and, thus, lower latency. %

In summary, we observe that Spark's throughput can be increased at the cost of increased latency and latency can be decreased at the cost of reduced throughput.
This means both metrics should always be considered in relation to each other and providing a single result value is problematic.
Nevertheless, we observe that with our benchmark, Spark achieves a throughput similar to the other frameworks only if a latency of a few minutes can be tolerated.
As we consider this as too long for most stream processing use cases~\cite{tudoran2016,Karimov2018}
and to reduce the space of experiment configurations, 
we limit the maximum number of records pulled per batch to 1\,000\,000 in the following throughput experiments. %

\subsection{Baseline Evaluation of Latency}\label{sec:evaluation:latency}

For our baseline latency evaluations, we generate a constant load of 90\,000 records per second as our baseline throughput experiments in \cref{sec:evaluation:throughput} showed that this is the maximum rate at which all frameworks are able to process data (see \cref{fig:barplot-baseline-stp}).

\begin{figure}
	\centering
	\includegraphics[width=\linewidth]{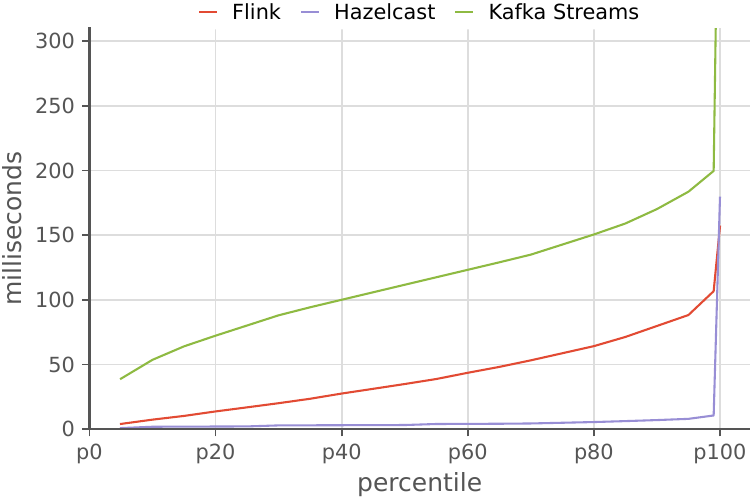}
	\caption{Quantile function of the median observed latency of Flink, Hazelcast and Kafka Streams.}
	\label{fig:quantileplot-baseline-lty}
\end{figure}

\Cref{fig:quantileplot-baseline-lty} shows a quantile function of the observed latency of Flink, Kafka Streams, and Hazelcast. We exclude Spark in the figure as it has a latency of over 10~seconds at every percentile. It can be seen that Hazelcast processes data with a very low latency of 8~milliseconds at the 95th percentile. The p95 latency of Flink (88~milliseconds) and Kafka Streams (183~milliseconds) are considerably higher but still by an order of magnitude lower than Spark's.

\subsection{Evaluation of Deployment Impact}\label{sec:evaluation:deployment}

In our baseline experiments, we deploy 9~instances of the stream processing framework, each limited to one virtual CPU core. Thus, all parallelization is happening on the level of Kubernetes Pods.
We compare this deployment against deploying 3~instances with each being limited to 3~cores. This introduces a second level of parallelization (i.e., within a single instance) while maintaining an overall parallelism of 9. We scale the assigned memory per instance proportionally to 12~GB.

\begin{figure}
	\begin{subfigure}[c]{\linewidth}
		\includegraphics[width=\linewidth]{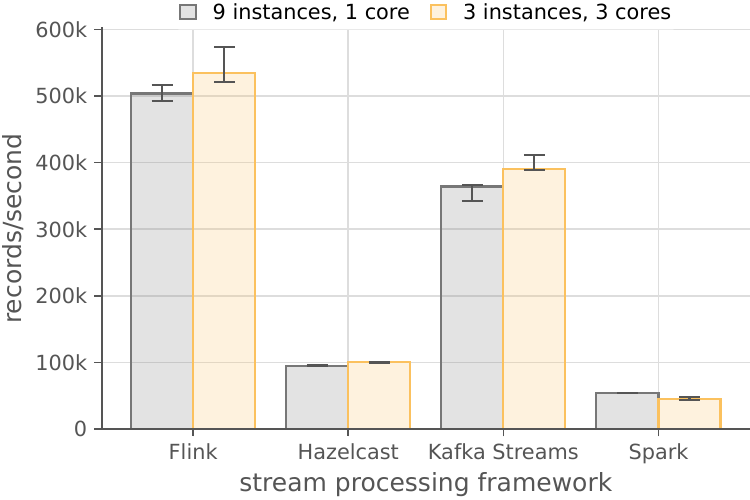}
		\caption{Ad-hoc throughput}
		\label{fig:barplot-multicore-atp}
	\end{subfigure}
	\begin{subfigure}[c]{\linewidth}
		\includegraphics[width=\linewidth]{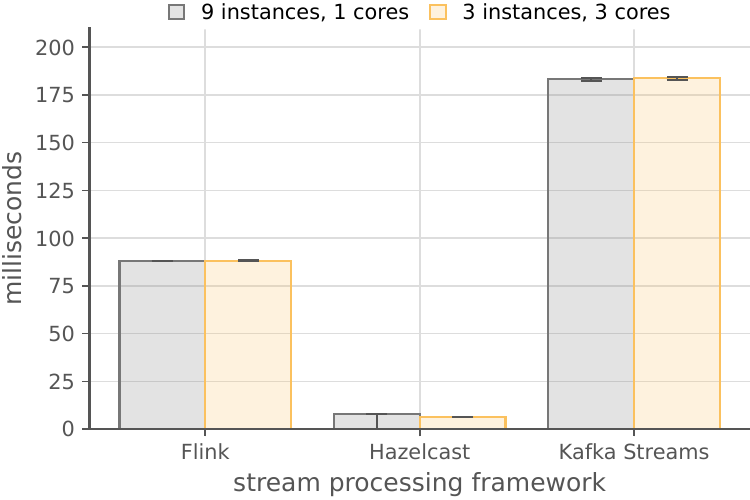}
		\caption{Latency at the 95th percentile}
		\label{fig:barplot-multicore-lty}
	\end{subfigure}
	\caption{Experimental results comparing a deployment with 9~instances and one core per instance with a deployment of 3~instances with 3~cores per deployment.}
	\label{fig:eval:multicore}
\end{figure}

Flink, Hazelcast, and Kafka Streams benefit to a small extent from higher per-instance parallelization as shown in \cref{fig:barplot-multicore-atp}. However, Spark's throughput decreases slightly.
\Cref{fig:barplot-multicore-lty} shows that the 3-node deployment has virtually no influence on processing latency for all frameworks.

\subsection{Evaluation of Record Size Impact}\label{sec:evaluation:recordsize}

In our baseline experiments, we generate data records of 1024-byte size. While we consider this as a realistic value for our studied industrial use case, there are certainly use cases that process records of other sizes. As related benchmarking studies often used considerably smaller records~\cite{vanDongen2020,Hesse2021,BDR2021}, we evaluate the performance with record sizes of 128~bytes, 256~bytes, and 512~bytes.

\begin{figure}
	\begin{subfigure}[c]{\linewidth}
		\includegraphics[width=\linewidth]{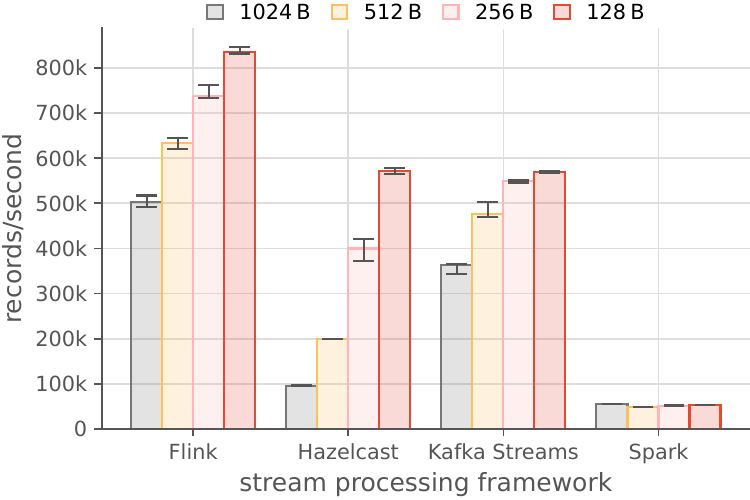}
		\caption{Ad-hoc throughput}
		\label{fig:barplot-recordsize-atp}
	\end{subfigure}
	\begin{subfigure}[c]{\linewidth}
		\includegraphics[width=\linewidth]{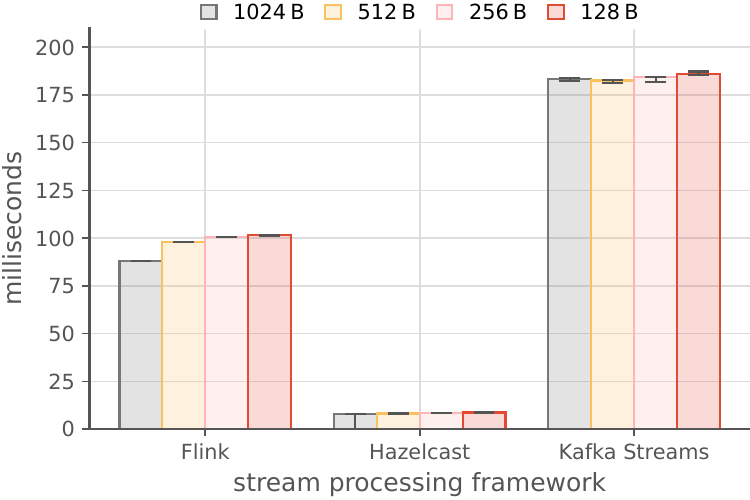}
		\caption{Latency at the 95th percentile}
		\label{fig:barplot-recordsize-lty}
	\end{subfigure}
	\caption{Experimental results comparing the impact of different record sizes.}
	\label{fig:eval:recordsize}
\end{figure}

\Cref{fig:barplot-recordsize-atp} shows that all frameworks can process more records as record sizes become smaller. It is interesting to see that in particular Hazelcast benefits from smaller record sizes.
While with 1024-byte records, Flink achieves $5.3\times$ higher and Kafka Streams $3.8\times$ higher throughput than Hazelcast, Hazelcast achieves 68\% of the throughput of Flink and approximately the same throughput as Kafka Streams with 128-byte records.
It is also worth noting that records smaller than 256 bytes have only a minimal impact on Kafka Streams’ throughput.

\cref{fig:barplot-recordsize-lty} shows that the record size only slightly affects the processing latency.
We can even see a small increase in latency with smaller records, which may seem counterintuitive. We expect this to be because data from Kafka is pulled in batches (limited by its size in bytes). Smaller records imply more records per batch and, thus, that data is potentially longer queued in Kafka before being pulled.

\subsection{Evaluation of Consumer Count Impact}\label{sec:evaluation:numconsumers}

ShuffleBench allows for configuring the number of real-time consumers. The number of consumers impacts the runtime of the matcher service as well as the number of state entries and, thus, the overall state size. Our baseline evaluation used 1~million consumers, which we now compare to 100\,000 consumers. Evaluations with significantly larger numbers would require more memory and are therefore not covered in the evaluation.

\begin{figure}
	\begin{subfigure}[c]{\linewidth}
		\includegraphics[width=\linewidth]{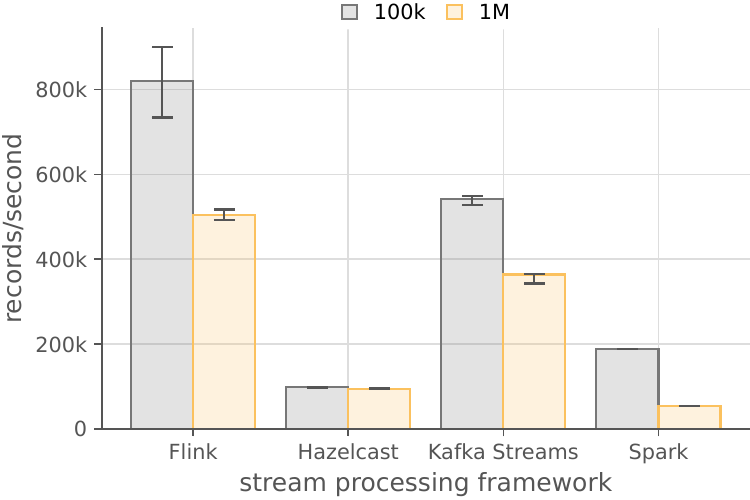}
		\caption{Ad-hoc throughput}
		\label{fig:barplot-numconsumers-atp}
	\end{subfigure}
	\begin{subfigure}[c]{\linewidth}
		\includegraphics[width=\linewidth]{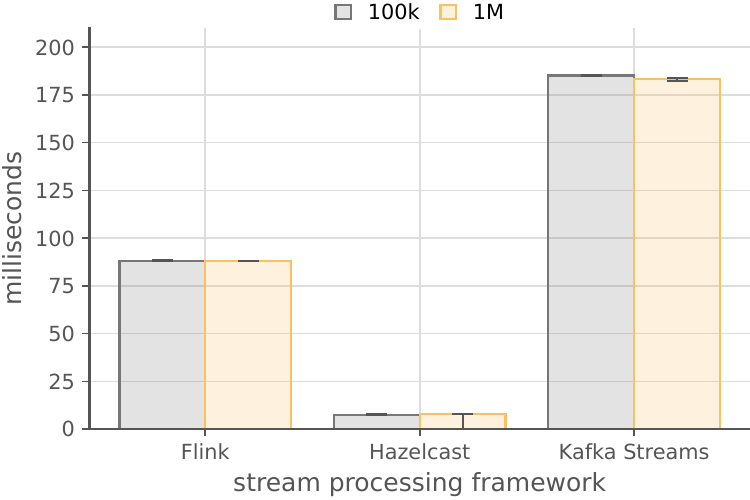}
		\caption{Latency at the 95th percentile}
		\label{fig:barplot-numconsumers-lty}
	\end{subfigure}
	\caption{Experimental results comparing 100\,000 with one million real-time consumers.}
	\label{fig:eval:numconsumers}
\end{figure}

\Cref{fig:barplot-numconsumers-atp} shows that with fewer real-time consumers, the throughput of Flink and Kafka Streams significantly increases, whereas we can see no change for Hazelcast. This indicates that the bottleneck of Hazelcast is unrelated to the number of consumers. %
With all frameworks, there is no change in the processing latency when using fewer consumers as shown in \cref{fig:barplot-numconsumers-lty}.

\subsection{Evaluation of Selectivity Impact}\label{sec:evaluation:selectivity}

The summed-up selectivity of all real-time consumes describes how much of the overall input data volume is forwarded to the \emph{groupBy} operation of ShuffleBench's dataflow architecture. Hence, it also determines the data volume that is re-distributed to a potentially different application instance. Besides a total selectivity of 20\% in our baseline evaluation, we also compare a 0\% and a 100\% total selectivity.

\begin{figure}
	\begin{subfigure}[c]{\linewidth}
		\includegraphics[width=\linewidth]{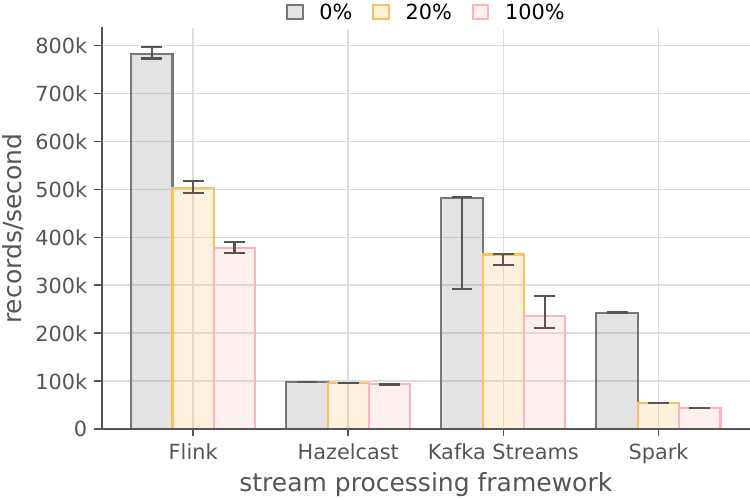}
		\caption{Ad-hoc throughput}
		\label{fig:barplot-selectivity-atp}
	\end{subfigure}
	\begin{subfigure}[c]{\linewidth}
		\includegraphics[width=\linewidth]{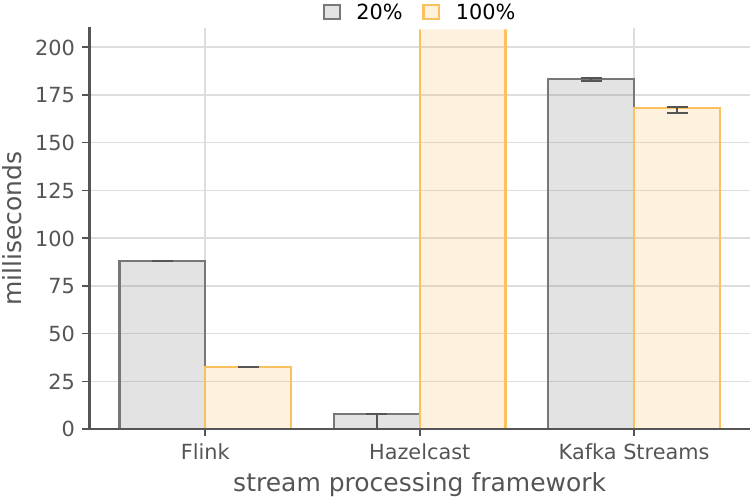}
		\caption{Latency at the 95th percentile}
		\label{fig:barplot-selectivity-lty}
	\end{subfigure}
	\caption{Experimental results comparing a total selectivity across all real-time consumers of 0\%, 20\%, and 100\%.}
	\label{fig:eval:selectivity}
\end{figure}

\Cref{fig:barplot-selectivity-atp} shows the throughput results for all frameworks. All frameworks except Hazelcast achieve a higher throughput with smaller selectivity. Hazelcast's results indicate that its bottleneck is neither due to the shuffling nor the stateful aggregation. Noticeable is that Spark's results throughput significantly reduces from 0\% selectivity to 20\% selectivity, but only slightly from 20\% selectivity to 100\% selectivity.
\Cref{fig:barplot-selectivity-lty} shows the 95th percentile processing latency of Flink, Hazelcast, and Kafka Streams for 20\% selectivity and 100\% selectivity.
There can be no latency results for 0\% selectivity, as no data is shuffled and, thus, no events are output.
Hazelcast shows a very high latency with 100\% selectivity as, in this case, it can not process the generated data volume anymore, which means that records are queuing up in Kafka.
Surprisingly, with higher selectivity, the latency of Flink and Kafka Streams decreases.
We expect this again to be due to buffering effects in the shuffling and in the output step.

\subsection{Discussion in the Context of Related Work}\label{sec:evaluation:discussion}

Across all our experiments, Flink processes data with the highest throughput, followed by Kafka Streams. Only for small records of 128-byte size, Hazelcast achieves a similar throughput to Kafka Streams.
On the other hand, Hazelcast processes data with about ten times lower latency compared to Flink, which in turn requires approximately half the time of Kafka Streams. In our experiments with Spark, a strong correlation between throughput and latency can be noted.  Engineers and operators can thus choose the right balance depending on the use case. To process data of a volume similar to that processed by other frameworks, however, latency in the order of several seconds up to minutes has to be tolerated.

Flink's superior performance for many use cases is also reported in the literature~\cite{Karimov2018,vanDongen2020,JSS2024}. The throughput achieved with Spark was not as high (compared to other frameworks) as reported in some related studies~\cite{vanDongen2020,Karakaya2017}. We believe that this is due to specific technical aspects of our benchmark and experimental design of avoiding configuration tuning (see Section~\ref{sec:evaluation:threats-to-validity}), which can be a requirement for Spark's execution in clusters~\cite{vanDongen2020}. In the future, we intend to further investigate Spark's configurations and their impact on performance.
Likewise, Hazelcast's low throughput is surprising and could not be observed in related work~\cite{JSS2024}. Whether this is only due to our larger record sizes or there are further particularities in our benchmark or experimental setup has to be further evaluated.

Hazelcast's low latency processing was also observed in the related literature~\cite{Gencer2021,Hesse2021} and is underpinned by its design~\cite{Gencer2021}.
Also, the strong correlation between throughput and latency in Spark is reported in the literature~\cite{SEAA2023,vanDongen2020}. Recent, experimental approaches in Spark for continuous processing\footnote{\url{https://spark.apache.org/docs/3.5.0/structured-streaming-programming-guide.html\#continuous-processing}} to reduce latency could be subject of an extended evaluation.

\subsection{Threats to Validity}\label{sec:evaluation:threats-to-validity}

Despite careful research design, there are threats and limitations to the validity of our experimental evaluation, which we report below.

\paragraph{Threats to Internal Validity}

We run all evaluations in a containerized environment on a public cloud platform to have a representative deployment. This means, however, that there are potentially many factors influencing the performance, which are out of our control~\cite{Leitner2016,Papadopoulos2021}.
We address these limitations to some extent by running individual experiments for a longer time, repeating them, and assessing their variability.
However, to reduce the benchmarking setup's complexity, we do not re-create the Kubernetes cluster between two experiments, which means that virtual machines are not re-provisioned before each experiment. We do not systematically run experiments at different times of the day or the week to keep the time required for these experiments manageable. Repeating certain experiments at different times, however, did not show noticeable deviations.
To further mitigate these limitations, we provide our benchmark as open-source software, allowing for independent replication of our study.

\paragraph{Threats to External Validity}

It is important to note that our results report on the performance of different stream processing frameworks for large-scale data shuffling use cases. Previous research has shown that benchmark results of one use case are not necessarily transferable to other use cases~\cite{JSS2024,SEAA2023,Pagliari2020}.
In \crefrange{sec:evaluation:deployment}{sec:evaluation:selectivity} we conduct throughput experiments with the less representative ad-hoc measurement method to reduce the overall time required for these experiments. \Cref{sec:evaluation:throughput} quantifies how much this method overestimates the realistic throughput.
We intentionally evaluate all frameworks primarily using their default configurations. This approach helps to avoid  bias resulting from different degrees of experience with the frameworks.
However, we can only draw limited conclusions about potential performance improvements that can be achieved through fine-tuning for specific scenarios.
For our experiments, we focused on a single kind of deployment (containers in Kubernetes) on a single cloud platform.
As we measure performance on a high level (macro-benchmarking) using cloud-native abstraction layers and by setting resource limits for the containers, we expect no significantly different results in other execution environments.

\section{Conclusions and Future Work}\label{sec:conclusions}

This paper introduces ShuffleBench, our proposal for a new stream processing benchmark for large-scale data shuffling operations.
Besides addressing a different use case than other stream processing benchmarks, ShuffleBench also overcomes several limitations of other benchmarks. %
Our benchmark design is based on requirements identified from a large cloud observability platform and established benchmarking metrics, methods, and techniques from the performance engineering research community.
With ShuffleBench, we aim to support and foster research on stream processing by providing a standardized method that researchers and practitioners can use to compare their implementations, algorithms, and configurations.
So far, we employed ShuffleBench to evaluate throughput and latency
of Flink, Hazelcast, Kafka Streams, and Spark.
Our evaluation provides the most recent benchmark results to the research community and serves as a starting point for researchers and practitioners to conduct further evaluations with ShuffleBench.

Besides growing a community around ShuffleBench, we plan to support additional qualities such as reliability.
In particular, an empirical investigation of the interconnection of throughput, latency, and fault-tolerance is highly demanded.
We are also currently in the process of supporting and evaluating non-uniformly distributed record sizes, selectivities, state sizes, and output rates to address additional industrial requirements.

\begin{acks}
We would like to thank the Johannes Kepler University Linz and Dynatrace for co-funding this research.
\end{acks}

\bibliographystyle{ACM-Reference-Format}
\bibliography{references}

\end{document}